\documentclass{PoS}

\title{Effects of dynamical quarks on the spectrum of the
Wilson Dirac operator}

\ShortTitle{Microscopic Wilson Dirac spectrum for $N_f=1$}

\author{Gernot Akemann\\
        Department of Mathematical Sciences \& BURSt Research Centre\\
Brunel University West London, Uxbridge UB8 3PH, United Kingdom\\
        E-mail: \email{mastgga@brunel.ac.uk}}
\author{\speaker{Poul H. Damgaard}\\
        The Niels Bohr International Academy and Discovery Center\\
        The Niels Bohr Institute,
        Blegdamsvej 17, DK-2100 Copenhagen, Denmark\\ 
        E-mail: \email{phdamg@nbi.dk}}
\author{Kim Splittorff\\
        The Niels Bohr Institute,
        Blegdamsvej 17, DK-2100 Copenhagen, Denmark\\
        E-mail: \email{split@nbi.dk}}
\author{Jac Verbaarschot\\
        State University of New York\\
        Department of Physics and Astronomy\\
        Stony Brook, NY 11794-3800, USA\\
        E-mail: \email{verbaarschot@cs.physics.sunysb.edu}}

\abstract{Effects of dynamical quarks on the microscopic spectrum
of the Wilson Dirac operator are analyzed by means of effective field
theory. We consider the distributions of the real modes of the 
Wilson Dirac operator as well as the spectrum of the Hermitian Wilson Dirac
operator, and work out the case of one flavor in all detail.
In contrast to the quenched case, the theory has a mild sign
problem that manifests itself by giving a spectral density that
is not positive definite as the spectral gap closes.}   

\FullConference{The XXVIII International Symposium on Lattice Field Theory, 
Lattice2010\\
		June 14-19, 2010\\
		Villasimius, Italy}

\usepackage{graphicx} 

\begin{document}
\newcommand{\Tr}{{\rm Tr}}
\newcommand{\Str}{{\rm STr}}
\newcommand{\hm}{\hat m}
\newcommand{\ha}{\hat a}
\newcommand{\hz}{\hat z}
\newcommand{\hx}{\hat x}

\section{Introduction}

The low-lying spectrum of the Dirac operator is a fascinating
subject, which also contains important physics. 
Recently, three of the present authors presented an analysis of
the quenched spectrum of the Wilson Dirac operator $D_W$
\cite{Damgaard:2010cz}. Focus was on the low-lying real modes of $D_W$ 
and the Hermitian counterpart $D_5 = \gamma_5(D_W +m)$. Effects of the
lattice spacing $a$ were taken into account to leading order. A
chiral Random Matrix Theory that encapsulates these leading order
terms was then established. As the size $N$ of the random matrices
goes to infinity, a scaling regime is reached where this
chiral Random Matrix Theory coincides with what one obtains from
the Wilson chiral Lagrangian to leading order in $a$. This
extends in a precise manner the universal Random Matrix Theory
results of continuum fermions \cite{SV,ADMN} to Wilson fermions
in the microscopic scaling regime. In particular,
the appropriate definition of an $\epsilon$-regime \cite{LS}
for the low-lying eigenvalues of the Wilson Dirac operator 
is identified \cite{Damgaard:2010cz}. The corresponding
spectrum away from the microscopic limit was first analyzed
at the mean field level by Sharpe in ref. \cite{Sharpe}. The Letter
\cite{Damgaard:2010cz} was very much motivated by that work and a wish
to understand in detail and at an analytical level 
some of the results of the lattice simulations in ref. \cite{Luscher}.

Here we report on a study of the effect of dynamical quarks on
these results. Because the case of two light flavors is significantly
more difficult in terms of computational complexity, we take here
the first step of unquenching by considering $N_f=1$. This
case is of interest in its own right because there are no Goldstone bosons
 and hence no chiral Lagrangian at our
disposal. Nevertheless, effective field theory can be used to
describe in a precise way the leading-order effects of Wilson terms in 
lattice gauge theory also in this case. By projecting onto sectors
of a fixed number $\nu$ of real modes (counted with the sign of
their chiralities, see below), we can also establish a chiral
Random Matrix Theory with exactly the same properties as the effective
field theory in the scaling limit. As for continuum fermions, 
the effective field theory in each fixed sector looks just like
the leading term in an $\epsilon$-regime counting of a chiral Lagrangian.
Yet there are no Goldstone
bosons and hence no way
to systematically introduce a full-fledged space-time dependent 
chiral Lagrangian which could
incorporate sub-leading effects of an associated $\epsilon$-expansion.

\section{The effective field theory}

Chiral symmetry for  QCD with just one flavor is broken explicitly 
due to the $U(1)$ anomaly, and there are no Goldstone bosons. 
As a consequence, we do not have the toolbox
of chiral Perturbation Theory available. Leutwyler and Smilga
\cite{LS} faced a similar situation when dealing with the spectrum
of the continuum Dirac operator, and we will here follow the same
line of reasoning. In the continuum, the leading effect of a quark
mass $m$ is proportional to the four-volume $V$. Because the 
logarithmic derivative yields the chiral condensate $\Sigma$, it follows
that the partition function must read $Z \sim \exp[m\Sigma V]$.
This term corresponds to 
$m\bar{\psi}\psi$ in the QCD Lagrangian.
For Wilson fermions, the Symanzik effective action has additional
operators $\sim a^2(\bar{\psi}\psi)^2$. Such terms give an
additional contribution to the free energy of order $a^2$ so that now
\begin{equation}
Z ~=~ \exp\left[m\Sigma V - 2W_8V a^2\right]
\end{equation}
where $W_8$ is a so far unknown constant. We have chosen the parametrization
so that this constant is naturally positive (the factor of 2 is for
later convenience). 
As argued in ref. \cite{Damgaard:2010cz} a positive sign
of $W_8$ is follows from 
from the Hermiticity properties of $D_W$.
We define the appropriate
$\epsilon$-regime here by requiring that 
$$
\hm ~\equiv~ m\Sigma V ~~~~~~~{\rm and}~~~~\hat{a}^2 ~\equiv~ a^2W_8V
$$
remain fixed as $V \to \infty$. This is the regime where there is 
competition between $m$ and $a^2$ effects and one can
imagine that a phase transition may occur. This turns out to be a
transition to the Aoki phase \cite{Aokiclassic} (see also \cite{Heller}).  
Other countings can also be considered \cite{Bar}, but they are not
of direct interest to us here.

In the continuum, a chiral rotation $\alpha$ shifts the vacuum
angle $\theta \to \theta + \alpha$. Noting that the $a^2$-term in the
effective action comes from operators $a^2(\bar{\psi}\psi)^2$, 
we define
\begin{equation}
Z(\theta) ~=~ \exp\left[m\cos(\theta)\Sigma V - 2W_8V a^2\cos(2\theta)\right],
\end{equation} 
and its Fourier components read: 
\begin{equation}
Z_{\nu} ~\equiv~ \int_{-\pi}^{\pi}\frac{d\theta}{2\pi}
e^{i\nu\theta}\exp\left[m\cos(\theta)\Sigma V - 2W_8V a^2\cos(2\theta)\right]
\end{equation}
Inverting this, we recover the original partition function as a sum
over each $Z_{\nu}$ after setting $\theta=0$:
\begin{equation}
Z(\theta=0) ~=~ \sum_{\nu=-\infty}^{\infty} ~Z_{\nu} .
\end{equation}
Let us now look
at each $Z_{\nu}$ separately. Interestingly, 
\begin{equation}
Z_{\nu} ~\equiv~ \int_{U(1)}dU~
\det(U)^{\nu}\exp\left[\frac{1}{2}m\Sigma V\Tr[U + U^{-1}] - 
W_8V a^2\Tr[U^2 + U^{-2}]\right] .
\end{equation}
This looks exactly like
the zero-momentum piece of the leading terms of a $U(1)$ chiral
Lagrangian for Wilson fermions \cite{ChPT}.  However, there
are no Goldstone bosons, and 
the $U(1)$ 'degree of freedom' results from 
the angular integration variable of the Fourier 
transform.

For general $N_f$
there would also be double-trace terms like $(\Tr[U + U^{-1}])^2$
and $(\Tr[U - U^{-1}])^2$, but in this $U(1)$ case such terms
just change the normalization of $W_8$ after use of elementary
trigonometric identities.
 
\section{Low-lying modes of the Wilson Dirac Operator}

\noindent
To get spectral information for the Wilson Dirac operator we need either

* ~ Pairs of extra species with opposite statistics (the graded method
\cite{DOTV}) or

* ~ Replicas \cite{DS,SV}.

\noindent
Here we use the graded method. We thus add a bosonic quark and a 
corresponding additional fermionic quark, both with appropriate
sources. When these sources are set equal to each other, the
two additional determinants exactly cancel each other. In this limit,
the partition
function of this graded theory therefore equals the partition
function of QCD with the original one flavor.

The graded method can be used in the effective field theory as well.
Additional Grassmann integrations truncate and trivially converge, but
care must be taken to ensure convergence of the {\em bosonic}
integrations. This problem has been solved in the context
of continuum fermions in ref. \cite{DOTV}. 
The graded partition function is
\begin{equation}
Z_{2|1}({\hat{\cal M}},{\hat{\cal Z}}) = \int_{Gl(2|1)} dU {\rm Sdet}(U)^\nu
e^{i\frac{1}{2}{\Str}({\hat{\cal M}}[U-U^{-1}])+i\frac{1}{2}{\Str}({\hat{\cal
    Z}}[U+U^{-1}])+ {\hat{a}}^2{{\Str}(U^2+U^{-2})}} .
\label{Z_2|1}
\end{equation}
The source terms are
\begin{eqnarray}
{\hat{\cal M}}=\left(\begin{array}{ccc} \hat{m}_f & 0 & 0 \\ 0 & \hat{m} & 
0 \\ 0 & 0 &
   \hat{m}' \end{array}\right)
\quad \nonumber
{\hat{\cal Z}}=\left(\begin{array}{ccc} \hat{z}_f & 0 & 0 \\ 0 & \hat{z} & 0 
\\ 0 & 0 &
    \hat{z}' \end{array}\right)
\end{eqnarray}
and when $\hat{m} = \hat{m}'$ and $\hat{z} = \hat{z}'$ a little miracle
occurs: the graded partition function becomes equal to the original
partition function of $N_f=1$. This follows from general principles,
but it arises in a highly non-trivial manner from the actual 
integrations of eq. (\ref{Z_2|1}). An explicit 
parametrization of the graded matrix $U$ has been provided in
ref. \cite{DOTV}:
\begin{eqnarray}
U &=& \left(\begin{array}{ccc} 
e^{it+iu}\cos(\theta) & ie^{it+i\phi}\sin(\theta) & 0 \\ 
ie^{it-i\phi}\sin(\theta) & e^{it-iu}\cos(\theta) & 0 \\ 
0 & 0 & e^{s}
  \end{array}\right)
\exp\left(\begin{array}{ccc}
0 & 0 & \alpha_1 \\ 
0 & 0 & \alpha_2 \\ 
\beta_1 & \beta_2 & 0 \end{array}\right)\nonumber
\end{eqnarray}
where $\theta,t,u\in[-\pi,\pi]$ and $\phi\in[0,\pi]$. The bosonic
degree of freedom $s$ is integrated over the real line, and
the $\alpha$'s and $\beta$'s are Grassmann variables.

A careful reader will have noticed the unusual form of (\ref{Z_2|1}).
Before extending the theory to the graded case, a rotation $U \to iU$
has been performed. In the original $U(1)$-integral this
simply shifts the angular variable
by $\pi/2$, while still integrating it over the full circle. Doing such
a rotation prior to extending the action to the graded case 
corresponds to a particular path of integration for the bosonic
variable $s$. It is the integration path
used in eq. (\ref{Z_2|1}) which corresponds to a non-Hermitian
(but $\gamma_5$-Hermitian) Wilson Dirac operator $D_W$.

The integrals in eq. (\ref{Z_2|1}) are tedious but doable. We have
performed the Grassmann integrations and one of the angular integrations
explicitly. The resulting expressions will be 
published elsewhere. Here we choose to present our results in a 
graphical manner.

While the Wilson Dirac Operator is not Hermitian, it is important
that it nevertheless retains $\gamma_5$-Hermiticity: $D_W^{\dagger}
= \gamma_5D_W\gamma_5$. Indeed, it is this property that ensures
Hermiticity of $D_5$. The spectrum of $D_W$ thus lies in the complex
plane, each non-real eigenvalue being matched by its complex conjugate partner.
To compute the spectrum of the (non-Hermitian) Wilson Dirac operator
by analytical means is slightly cumbersome because of this. However, $D_W$
also has a certain number of eigenvalues sitting on the real line.
The distribution of 
the chiralities of the corresponding states over the Dirac spectrum 
is much easier to compute. To this end, 
let us define a resolvent (and put $\hz=\hz'=0$)
\begin{equation}
\Sigma^\nu(\hm_f,\hm) ~\equiv~ 
\lim_{{\hm}'\to \hm} \frac{\partial}{\partial\hm} 
\ln Z^\nu_{2|1}(\hm_f,\hm,{\hm}')  .
\end{equation}
The discontinuity across the real line gives us 
the  distribution of the chiralities over the Dirac spectrum 
\begin{equation}
\rho^\nu_{\rm \chi}(\hat\zeta) 
 \equiv  \sum_{k, \zeta_k \in {\mathcal R}}
\delta(\hat \zeta -\hat \zeta_k) \ \chi_k  \ = \
\frac{1}{\pi}{\rm Im}[\Sigma^\nu(\hm_f,\hat\zeta)] 
\end{equation}
with the chirality 
$ \chi_k = {\rm sign }(\langle k| \gamma_5 |k \rangle )$.

\begin{figure}[htb]
\vspace{0.6cm}
\begin{center}
\includegraphics[width=10cm]{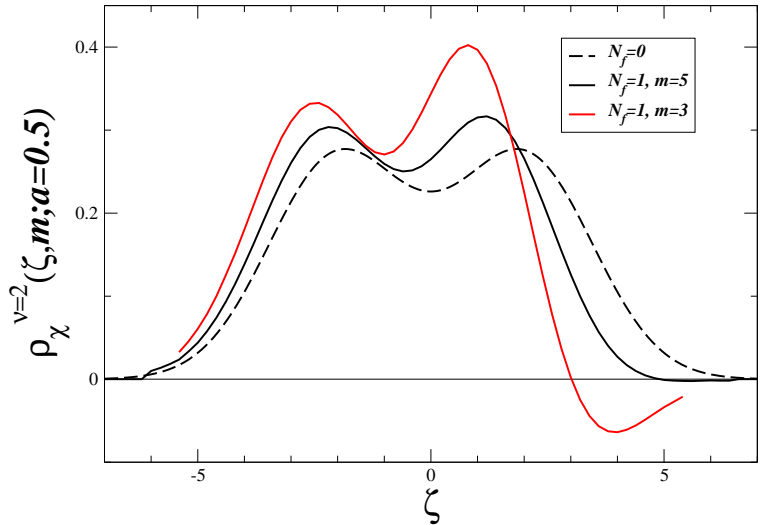}
\caption{The density of chiralities of the Wilson Dirac operator.}
\end{center}
\end{figure}
The integral over this distribution is normalized to $\nu$,
\begin{equation}
\int_{-\infty}^{\infty}d\hat{\zeta} ~\rho^\nu_{\chi}(\hat\zeta) ~=~ \nu.
\end{equation}
The index $\nu$ counts chiralities of the real modes of $D_W$ in
$Z_\nu$: $\nu = \sum_n \chi_n$ 
where $n$ runs over all real modes. 
In the limit of small $a$ the probability
of finding configurations with real modes that have chiralities of
different signs vanishes. In that limit $\nu$ is simply the number of
real modes.
The non-positivity of the density of real
modes is unrelated to this: a change of sign occurs at $\hat{\zeta}
= \hat{m}$. Only when $\hat{m}$ is on the order of or less than 
$8\hat{a}^2 $ does this have significance in the
density since otherwise the density is very small anyway. 
We show an example of the distribution of the chiralities 
over the real modes in fig. 1.
  
We now wish to compute the spectrum of the
Hermitian Wilson Dirac operator $D_5 = \gamma_5(D_W+m)$. To that end,
introduce the new resolvent
\begin{equation}
G^\nu(\hz,\hm) ~\equiv~ 
\lim_{{\hz}'\to \hz} \frac{\partial}{\partial\hz} 
\ln Z^\nu_{2|1}(\hm,\hm,\hm,0,\hz,{\hz}') ~=~ 
\left\langle{\rm Tr}\left(\frac{1}{D_5+\hz}\right)\right\rangle
\end{equation}
and take the discontinuity across the real line. This gives us
the spectral density of $D_5$:
\begin{equation}
\rho^\nu_5(\hx) ~=~ \frac{1}{\pi}{\rm Im}[G^\nu(\hx,\hm)] .
\end{equation}

Let us first consider the spectrum corresponding to $\nu=0$. 
In fig. 2 we show the density for fixed $\hat{m}=5$ and various
values of $\hat{a}^2$. When $\hat{a}$ is small, a gap clearly opens
up around $\pm \hat{m}$, as it should. The spectrum of $D_5$
then approaches the standard spectrum of the continuum Dirac
operator with one massive fermion \cite{DN}, up to a trivial change of
variables. In contrast to the quenched spectrum, 
the microscopic spectrum of $D_5$ in 
this $N_f=1$ theory always has a zero at the origin. This is clearly visible
in fig. 2.

\begin{figure}[t]
\vspace{0.6cm}
\begin{center}
\includegraphics[width=10cm]{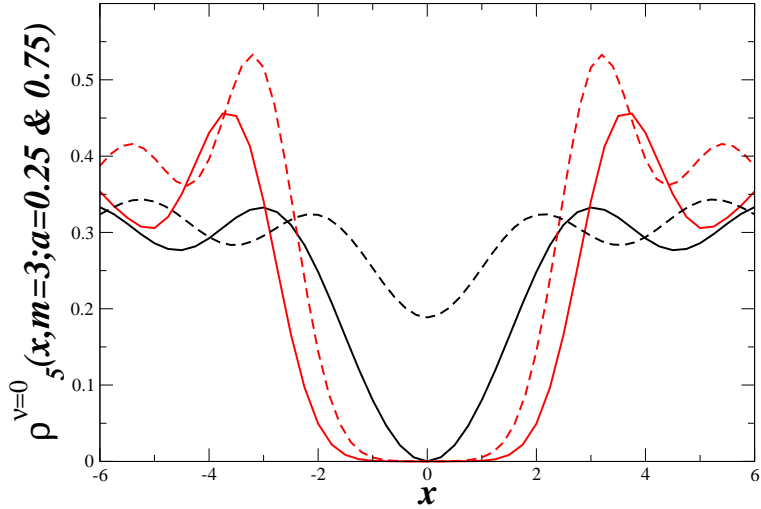}
\vspace{0.2cm}
\caption{Spectral density of the Hermitian Wilson Dirac operator for
$\nu=0$. (Dashed lines $N_f=0$.)}
\end{center}
\end{figure}

There are other differences with the quenched spectrum. Because of the
real modes, the spectrum of $D_5$ can change sign in the $N_f = 1$ theory.
A negative density simply corrersponds to
a theory with a sign problem: the Boltzmann weight in the path integral
is not positive definite. The existence of a negative density is thus
a potential problem for numerical simulations. Fortunately the sign
problem in this theory is mild: it is only signifact
in the small-$m$ limit, and it can be postponed by 
going to smaller lattice spacings $a$. We illustrate this phenomenon
in fig. 3, where we consider a case with
$\nu=1$. The analytical
understanding we can provide here should be valuable
for numerical simulations.

\begin{figure}[htb]
\vspace{0.6cm}
\begin{center}
\includegraphics[width=10cm]{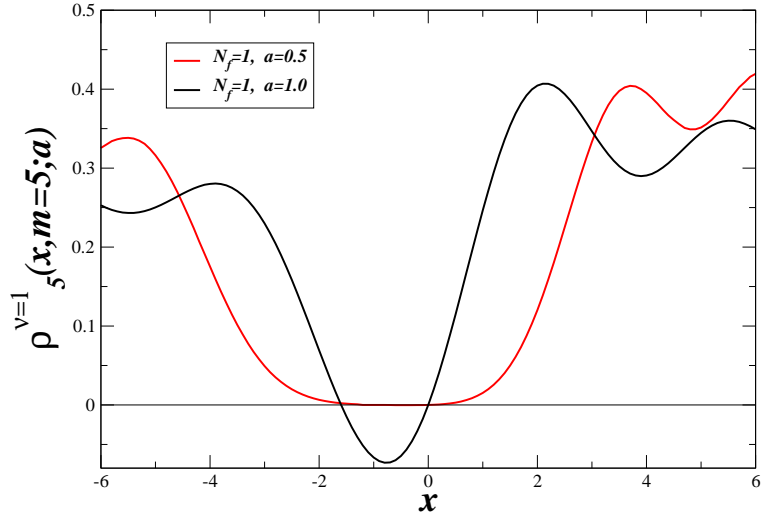}
\vspace{0.2cm}
\caption{Same as fig.2, but now for $\nu=1$. The spectral density is no
longer positive.}
\end{center}
\end{figure}

\section{Conclusions}

We have presented an explicit computation of
the microscopic  eigenvalue distributions of the Wilson Dirac operator, the
real modes of this operator, and the eigenvalues of the
Hermitian Wilson Dirac operator. We have  
focused on effects that most clearly distinguish
a theory with dynamical quarks from the quenched counterpart
\cite{Damgaard:2010cz}. A sum over the index 
$\nu$ can be done straightforwardly.
This will be presented elsewhere.

{\sc Acknowledgment:}~ 
We would like to thank many participants of the Lattice 2010 Symposium and
the TH-Institute ``Future directions in lattice
gauge theory - LGT10'' at CERN for discussions. Two of us (GA and JV) would
like to thank the Niels Bohr Foundation for support and the Niels
Bohr Institute and the Niels Bohr International Academy for its
warm hospitality. 
This work was supported by U.S. DOE Grant No. 
DE-FG-88ER40388 (JV) and the 
Danish Natural Science Research Council (KS).

\end{document}